# Direct Interband Light Absorption in Strongly Prolated Ellipsoidal Quantum Dots' Ensemble

K. G. Dvoyan · D. B. Hayrapetyan ·
E. M. Kazaryan

**Abstract** Within the framework of adiabatic approximation, the energy levels and direct interband light absorption in a strongly prolated ellipsoidal quantum dot are studied. Analytical expressions for the particle energy spectrum and absorption threshold frequencies in three regimes of quantization are obtained. Selection rules for quantum transitions are revealed. Absorption edge and absorption coefficient for three regimes of size quantization (SQ) are also considered. To facilitate the comparison of obtained results with the probable experimental data, size dispersion distribution of growing quantum dots by the small semiaxe in the regimes of strong and weak SQ by two experimentally realizing distribution functions have been taken into account. Distribution functions of Lifshits–Slezov and Gaussian have been considered.

**Keywords** Ellipsoidal quantum dot ·
Interband light absorption · Selection rules ·
Quantum dot ensemble

## Introduction

Development of the novel growth techniques, such as the Stranski–Krastanov epitaxial method etc., makes possible to grow semiconductor quantum dots (QDs) of various shapes and sizes [1–3]. As is known, the energy spectrum of charge carriers in QDs is completely quantized and resembles the energy spectrum of atoms (artificial atoms) [4]. In recent years, many theoretical and experimental works have evolved, where ellipsoidal, pyramidal, cylindrical, and lens-shaped QDs were considered [5–13]. As a result of diffusion, the confining potential, formed during the growth process, in most cases can be approximated with a high accuracy by a parabolic potential. However, an effective parabolic potential may arise in a QD in view of features of its external shape [14]. In particular, the case in point is a QD having the shape of a strongly prolated ellipsoid of revolution [15].

Investigations of the optical absorption spectrum of various semiconductor structures are a powerful tool for determination of many characteristics of these systems: forbidden band gaps, effective masses of electrons and holes, their mobilities, dielectric permittivities, etc. There are many works devoted to the theoretical and experimental study of the optical absorption both in massive semiconductors and size-quantized systems. The presence of size quantization (SQ) essentially influences the absorption mechanism. In fact, the formation of new energy levels of the SQ makes possible new interlevel transitions.

In this paper, the electron states and direct interband absorption of light in a strongly prolated ellipsoidal QD (SPEQD) at three regimes of SQ is considered. Absorption edge and absorption coefficient for three regimes of SQ are also considered. To facilitate the comparison of obtained results with the probable experimental data, size dispersion distribution of growing QDs by the small semiaxe in the regimes of strong and weak SQ by two experimentally realizing distribution functions have been taken into account. Distribution function of Lifshits–Slezov has been

K. G. Dvoyan (✉) · D. B. Hayrapetyan · E. M. Kazaryan
Department of Applied Physics and Engineering,
Russian-Armenian State University, 123 Hovsep Emin Street,
Yerevan 0051, Armenia
e-mail: dvoyan@gmail.com dhayrap@web.am

D. B. Hayrapetyan
Department of Physics, State Engineering University
of Armenia, 105 Terian Street, Yerevan 0009, Armenia
e-mail: dhayrap@web.am

considered in the first model and distribution function of Gauss has been considered in the second case.

## Theory

Regime of Strong Size Quantization

Consider the motion of a particle (electron, hole) in an SPEQD (see Fig. 1). The particle potential energy in cylindrical coordinates can be written as

$$U(\rho,\varphi,Z) = \begin{cases} 0, & \frac{\rho^2}{a_1^2}+\frac{Z^2}{c_1^2} \leq 1 \\ \infty, & \frac{\rho^2}{a_1^2}+\frac{Z^2}{c_1^2} > 1 \end{cases}, \quad a_1 <\!< c_1, \quad (1)$$

where $a_1$ and $c_1$ are the small and large semiaxes of the SPEQD, respectively.

In the regime of strong SQ, the energy of the Coulomb interaction between the electron and hole is much less than the energy caused by the SQ contribution. In this approximation, the Coulomb interaction can be neglected. Then, the problem is reduced to the determination of separate energy states of the electron and hole. It follows from the geometrical form of a QD that the particle motion along the radial direction occurs more rapidly than along the Z-direction. This allows one to use the adiabatic approximation [16]. The Hamiltonian of the system in the cylindrical coordinates has the form

$$\hat{H} = -\frac{\hbar^2}{2\mu_p}\left[\frac{\partial^2}{\partial\rho^2}+\frac{1}{\rho}\frac{\partial}{\partial\rho}+\frac{1}{\rho^2}\frac{\partial^2}{\partial\varphi^2}\right]-\frac{\hbar^2}{2\mu_p}\frac{\partial^2}{\partial Z^2} \\ + U(\rho,\varphi,Z), \quad (2)$$

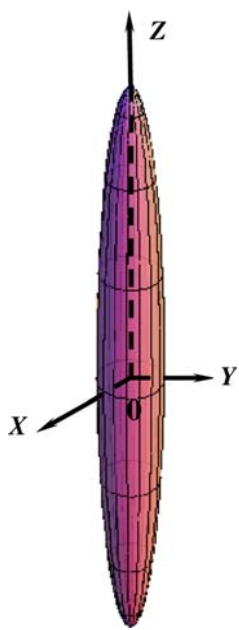

**Fig. 1** Strongly prolated ellipsoidal quantum dot

and it can be represented as a sum of the Hamiltonians for the "fast" ($\hat{H}_1$) and "slow" ($\hat{H}_2$) subsystems in dimensionless quantities:

$$\hat{H} = \hat{H}_1 + \hat{H}_2 + U(r,\varphi,z), \quad (3)$$

where

$$\hat{H}_1 = -\left[\frac{\partial^2}{\partial r^2}+\frac{1}{r}\frac{\partial}{\partial r}+\frac{1}{r^2}\frac{\partial^2}{\partial\varphi^2}\right], \quad (4)$$

$$\hat{H}_2 = -\frac{\partial^2}{\partial z^2}. \quad (5)$$

Here, $\hat{H}=\frac{\hat{H}}{E_R}$, $r=\frac{\rho}{a_B}$, and $z=\frac{Z}{a_B}$; $\mu_p$ is the effective mass of the particle, $E_R = \frac{\hbar^2}{2\mu_p a_B^2}$ is the effective Rydberg energy, $a_B = \frac{\kappa\hbar^2}{\mu_p e^2}$ is the effective Bohr radius of the particle, $e$ is the particle charge, and $\kappa$ is the dielectric constant. The wave function can be sought in the form

$$\psi(r,\varphi,z) = e^{im\varphi}R(r;z)\chi(z). \quad (6)$$

At a fixed value of the slow subsystem z-coordinate, the particle motion is localized in a two-dimensional potential well with the effective variable width

$$L(z) = a\sqrt{1-\frac{z^2}{c^2}}, \quad (7)$$

where $a=\frac{a_1}{a_B}$ and $c=\frac{c_1}{a_B}$. From the fast subsystem Schrödinger equation, we obtain particle energy spectrum

$$\varepsilon_1(z) = \frac{\alpha_{n+1,m}^2}{L^2(z)}, n=0,1,2\ldots, m=0,\pm 1,\pm 2\ldots, \quad (8)$$

where $\alpha_{n+1,m}$ are the zeros of the first-kind $J_m(r)$ Bessel function. For the lower levels of the spectrum, the particle is mainly localized in the region $|z| <\!< a$. Based on this, we expand $\varepsilon_1(z)$ into a series

$$\varepsilon_1(z) \approx \varepsilon_0 + \omega_0^2 z^2, \quad (9)$$

where $\varepsilon_0 = \frac{\alpha_{n+1,m}^2}{a^2}$, $\omega_0 = \frac{\alpha_{n+1,m}}{ac}$. Expression (9) is the effective potential entering the Schrödinger equation of the slow subsystem. For the total energy of the system, one derives

$$\varepsilon = \varepsilon_0 + 2\omega_0\left(N+\frac{1}{2}\right), N=0,1,2\ldots. \quad (10)$$

Regime of Intermediate Size Quantization

For this regime, one should consider the electron–hole interaction. It is evident that in view of SQ such an interaction is clearly exhibited only in the Z-direction. Therefore, we restrict ourselves to the case of a one-dimensional exciton. It is clear that in this SQ regime, the energy of electron motion predominates over the energy of heavy-hole motion (from the condition $\mu_e <\!< \mu_h$). Based on the above, the electronic potential acted on the hole can be averaged over the electron motion and written as

$$\tilde{V}_{n,m,N}(Z;r) = -\frac{e^2}{\kappa} \int_{-c_1}^{c_1} \frac{|\Psi_{n,m,N}(Z';r)|^2}{|Z-Z'|} dZ', \quad (11)$$

where $\Psi_{n,m,N}(Z')$ is the electron wave function. The condition $a_h \ll a_1$ allows us to expand potential (11) into a series near the point $Z = 0$, where $a_h = \frac{\kappa\hbar^2}{\mu_h e^2}$ is the effective Bohr radius of the hole. Finally we obtain for expression (11) in dimensionless quantities

$$V_{100}(r) = \alpha + \beta^2 r^2, \quad (12)$$

where the following notations are introduced:

$$V_{100}(r) = \frac{\tilde{V}_{100}(\rho)}{E_R}, \quad \alpha = \frac{3}{4}\frac{1}{c}\left(1 + \ln\frac{\alpha_{10}c}{a}\right)\sqrt{\frac{\alpha_{10}c}{a}},$$
$$\beta^2 = \frac{3}{4}\frac{\alpha_{10}}{a}\left(\ln\frac{\alpha_{10}c}{a} - 1\right)\sqrt{\frac{\alpha_{10}c}{a}}. \quad (13)$$

The hole wave function and energy are determined from the Schrödinger equation with averaged potential (12). After simple transformations, for the hole energy spectrum in dimensionless quantities, we derive

$$\varepsilon = \varepsilon' + \alpha + 2\sqrt{\omega'^2 + \beta^2}(N' + 1/2), N' = 0,1,2,\ldots, \quad (14)$$

where $\varepsilon' = \frac{\alpha_{n'+1,m'}^2}{a^2}$, $\omega' = \frac{\alpha_{n'+1,m'}}{ac}$.

Regime of Weak Size Quantization

In this case, the system's energy is caused mainly by the electron–hole Coulomb interaction. In other words, we consider the motion of an exciton as a whole in SPEQD. Then, the wave function of the system can be represented as

$$f(\vec{r}_e, \vec{r}_h) = \varphi(\vec{r})\Phi_{n_r,l,m}(\vec{R}), \quad (15)$$

where $\vec{r} = \vec{r}_e - \vec{r}_h$, $\vec{R} = \frac{\mu_e \vec{r}_e + \mu_h \vec{r}_h}{\mu_e + \mu_h}$. Here, $\varphi(\vec{r})$ describes the relative motion of the electron and hole, while $\Phi_{n,l,m}(\vec{R})$ describes the motion of the center of gravity of the exciton. The Hamiltonian of the system is written as

$$\hat{H} = -\frac{\hbar^2}{2M}\Delta_{\vec{R}}^2 - \frac{\hbar^2}{2\mu}\Delta_{\vec{r}}^2 - \frac{e^2}{\kappa r}, \quad (16)$$

where $M = \mu_e + \mu_h$, $\mu = \frac{\mu_e \mu_h}{\mu_e + \mu_h}$. For the exciton center of mass energy, one obtains the result analogous to formula (10), but by the mass $\mu_p$ is meant the exciton mass $M$. For the energy spectrum of the exciton's relative motion, we have in dimensionless quantities

$$\varepsilon_{ex} = \frac{E_{ex}}{E_R} = \frac{\mu}{M}\frac{1}{k^2}, k = 1, 2, \ldots. \quad (17)$$

Finally for the total energy, we derive

$$\varepsilon = \varepsilon_0 + 2\omega_0\left(N + \frac{1}{2}\right) - \varepsilon_{ex}, N = 0, 1, 2\ldots, \quad (18)$$

where $\varepsilon_0 = \frac{\alpha_{n+1,m}^2}{a^2}$, $\omega_0 = \frac{\alpha_{n+1,m}}{ac}$.

**Direct Interband Light Absorption**

Let us proceed to consideration of the direct interband absorption of light in an SPEQD in the regime of strong SQ. Consider the case of a heavy hole when $\mu_e \ll \mu_h$. The absorption coefficient is defined by the expression [18]

$$K = A\sum_{\nu,\nu'}\left|\int \Psi_\nu^e \Psi_{\nu'}^h d\vec{r}\right|^2 \delta(\hbar\Omega - E_g - E_\nu^e - E_{\nu'}^h), \quad (19)$$

where $\nu$ and $\nu'$ are the sets of quantum numbers corresponding to the electron and heavy hole, $E_g$ the forbidden band gap of a massive semiconductor, $\Omega$ the incident light frequency, $A$ a quantity proportional to the square of modulus of the matrix element of the dipole moment taken over the Bloch functions [17]. Finally, in the regime of strong SQ, for the quantity $K$ and the absorption threshold, we obtain

$$K = A\sum_{n,m,N}\left(\frac{J_{1+m}(\alpha_{n+1,m})}{J_{1-m}(\alpha_{n+1,m})}\right)^2 \delta(\hbar\Omega - E_g - E_e - E_h), \quad (20)$$

$$W_{000} = 1 + \alpha_{10}^2 \frac{d^2}{a_1^2} + \alpha_{10}\frac{d^2}{a_1 c_1}. \quad (21)$$

here $W_{000} = \frac{\hbar\Omega_{000}}{E_g}$ and $d = \frac{\hbar}{\sqrt{2\mu E_g}}$. Formula (21) characterizes the dependence of the effective forbidden band gap on the semiaxes $a_1$ and $c_1$. With increasing semiaxes, the absorption threshold decreases, but the dependence on the small semiaxis becomes stronger. Consider now the selection rules for transitions between the levels with different quantum numbers. For the magnetic quantum number, the transitions between the levels with $m = -m'$ are allowed, while for the quantum number of the fast subsystem the transitions with $n = n'$. For the oscillatory quantum number, the transitions for the levels with $N = N'$ are allowed. Note that the analytical form of expression (20) is given with allowance for the above-mentioned selection rules.

We proceed to consideration of the direct interband absorption of light in SPEQD in the regime of intermediate SQ. In this case, the consideration of the electron–hole interaction leads to the fact that in the spectrum of the interband optical absorption each line corresponding to given values of $\nu$ transforms into a set of closely spaced lines corresponding to different values of $v'$. In this regime of SQ, the absorption coefficient has the form

$$K = A \sum_{v,v'} \left| \int \Psi(\vec{r}_e, \vec{r}_h) \delta(\vec{r}_e - \vec{r}_h) d\vec{r}_e d\vec{r}_h \right|^2$$
$$\delta(\hbar\Omega - E_g - E_v^e - E_{v'}^h). \quad (22)$$

Finally we derive for the absorption coefficient and absorption threshold

$$K = A \sum_{\substack{m,n \\ N,N'}} \left| I_{N,N'}^{m,n} \right|^2 \delta(\hbar\Omega - E_g - E_v^e - E_{v'}^h), \quad (23)$$

$$W_{000} = 1 + \alpha_{10}^2 \frac{d^2}{a_1^2} + \alpha_{10} \frac{\mu}{\mu_e} \frac{d^2}{a_1 c_1} + \frac{\alpha E_h}{E_g} + \frac{E_h}{E_g} \sqrt{\omega'^2 + \beta^2}. \quad (24)$$

Here, $I_{N,N'}^{m,n}$ is an integral, which is calculated numerically, $W_{000} = \frac{\hbar\Omega_{000}}{E_g}$, and $d = \frac{\hbar}{\sqrt{2\mu E_g}}$. In this case, the transitions between the levels with $m = -m'$ and $n = n'$ are allowed. It should also be noticed that taking into account the effective one-dimensional Coulomb interaction leads to the destruction of the previous symmetry of the task and to the full removal of selection rules for the oscillatory quantum number $N$.

Let us consider the direct interband absorption of light in the regime of weak SQ. Taking into account the localization of an exciton in a relatively small vicinity of the QD center, for the absorption coefficient one can write the expression

$$K = A \sum_{n,n_r,l,m} |\varphi(0)|^2 \left| \int \Phi_{n,n_r,m}(\vec{R}) d\vec{R} \right|^2 \delta(\hbar\Omega - E_g - E) \quad (25)$$

where $E$ is the energy (18) in dimensional quantities. It should be noted that $\varphi(0) \neq 0$ only for the ground state when $l = m = 0$ ($l$ is the orbital quantum number). Finally, in the regime of weak SQ, we get for the absorption coefficient and absorption threshold the expressions

$$K = A \sum_{n_R,N_R} \left| J_{N_R}^{n_R} \right|^2 \delta(\hbar\Omega - E_g - E), \quad (26)$$

$$W_{1000} = 1 + \alpha_{10}^2 \frac{h^2}{a_1^2} + \alpha_{10} \frac{h^2}{a_1 c_1} - \frac{h^2}{a_{ex}^\mu a_{ex}^M}. \quad (27)$$

Here, $J_{N_R}^{n_R}$ denotes an integral, which is calculated numerically, $W_{1000} = \frac{\hbar\Omega_{1000}}{E_g}$, $h = \frac{\hbar}{\sqrt{2ME_g}}$, $a_{ex}^\mu = \frac{\kappa\hbar^2}{\mu e^2}$, and $a_{ex}^M = \frac{\kappa\hbar^2}{Me^2}$. The most important feature of this case is the fact that with changing semiaxes of the SPEQD the excitonic level shift is determined by the total mass of the exciton.

## Direct Interband Light Absorption with Account of Dispersion of QDs Geometrical Sizes

So far we have studied the absorption of a system consisting of semiconductor QDs having identical dimensions. For comparison of the obtained results with experimental data, one has to take into account the random character of SPEQD dimensions (or half-axis) obtained in the growth process. The absorption coefficient should be multiplied by concentration of QDs. Instead of distinct absorption lines, account of size dispersion will give a series of fuzzy maximums. In the first model, we use the Lifshits–Slezov distribution function [19]:

$$P(u) = \begin{cases} \frac{3^4 eu^2 \exp((-1/(1-2u/3)))}{2^{5/3}(u+3)^{7/3}(3/2-u)^{11/3}} & u < 3/2 \\ 0, & u > 3/2 \end{cases}, \quad u = \frac{a}{\bar{a}} = \frac{a_1}{\bar{a}_1}, \quad (28)$$

where $\bar{a}$ is some average value of the half-axis. In the second model, the Gaussian distribution function is used (see e.g., [20]):

$$P(u) = A e^{-\frac{(u-1)^2}{\sigma/\bar{a}}}. \quad (29)$$

In the case of strong SQ with account of general size distribution function $P(u)$, we obtain for the absorption coefficient corresponding formula:

$$K = A \sum_{m,n,N} \frac{(J_{1+m}(\alpha_{n+1,m})/J_{1-m}(\alpha_{n+1,m}))^2}{E_g \sqrt{\lambda_3^2 + 4\lambda_1\lambda_2}}$$
$$\times \left( \frac{2\lambda_2}{\sqrt{\lambda_3^2 + 4\lambda_1\lambda_2} - \lambda_3} \right)^2 P\left( \frac{2\lambda_2}{\sqrt{\lambda_3^2 + 4\lambda_1\lambda_2} - \lambda_3} \right), \quad (30)$$

where $\lambda_1 = \frac{\hbar\Omega - E_g}{E_g}$, $\lambda_2 = \alpha_{n+1,m}^2 \left(\frac{d_\mu}{\bar{a}_1}\right)^2 \left(1 + \frac{\mu_e}{\mu_h}\right)$, and $\lambda_3 = \alpha_{n+1,m} \frac{d_\mu^2}{\bar{a}_1 c_1}(N+1)\left(1 + \frac{\mu_e}{\mu_h}\right)$. In the case of weak SQ with account of general size distribution function $P(u)$, we obtain for the absorption coefficient corresponding formula:

$$K = A \sum_{n,n_r} \left| J_{N_R}^{n_R} \right|^2 \frac{1}{E_g} \frac{1}{\sqrt{\lambda_3^2 + 4\lambda_1\lambda_2}}$$
$$\times \left( \frac{2\lambda_2}{\sqrt{\lambda_3^2 + 4\lambda_1\lambda_2} - \lambda_3} \right)^2 P\left( \frac{2\lambda_2}{\sqrt{\lambda_3^2 + 4\lambda_1\lambda_2} - \lambda_3} \right), \quad (31)$$

where $\lambda_1 = \frac{\hbar\Omega - E_g}{E_g} + \frac{d_\mu^2}{a_{ex}^2} \frac{M}{\mu} \frac{1}{q^2}$, $\lambda_2 = \alpha_{n+1,m}^2 \left(\frac{d_\mu}{\bar{a}_1}\right)^2$, $\lambda_3 = \alpha_{n+1,m} \frac{d_\mu^2}{\bar{a}_1 c_1}(N+1)$, $a_{ex} = \frac{\kappa\hbar^2}{\mu e^2}$.

## Discussion

As is seen from formula (10), the energy spectrum of CCs in SPEQD is equidistant. This result is related only to the lower levels of the spectrum. Numerical calculations for the case of strong SQ were performed for a *GaAs* QD with the following parameters: $\mu_e = 0.067 m_e$, $\mu_e = 0.12\mu_h$, $\kappa = 13.8$, $E_R = 5.275$ meV, $a_e = 104$ Å and $a_h = 15$ Å are the effective Bohr radii of the electron and hole, $E_g = 1.43$ eV is the forbidden band gap of a massive semiconductor. In the strong SQ regime, the frequency of transition between the equidistant levels (for the value $n = 0$), at fixed values $a_1 = 0.5a_e$ and $c_1 = 2.5a_e$, is equal to $\omega_{00} = 3.32 \times 10^{13}$ s$^{-1}$, which corresponds to the infrared region of the spectrum. For the same values of quantum numbers, but with the values $a_1 = 0.4a_e$ and $c_1 = 2a_e$, we obtain $\omega_{10} = 5.19 \times 10^{13}$ s$^{-1}$, which is half as much again as the preceding case. As is seen from formula (10), with increasing semiaxes the particle energy is lowered. Note that this energy is more "sensitive" to changes of the small semiaxis, which is a consequence of the higher contribution of SQ into the particle energy in the direction of the axis of ellipsoid revolution. With increasing semiaxes, the energy levels come closer together, but remain equidistant.

Figures 2 and 3 present the dependences of the absorption threshold on the small and large semiaxes of the SPEQD, respectively. With decreasing semiaxes, the absorption threshold increases, which is a consequence of the higher contribution of SQ (the "effective" forbidden band gap increases). As is seen from the plots, the change in the absorption threshold is larger in the dependence on the small semiaxis of the SPEQD.

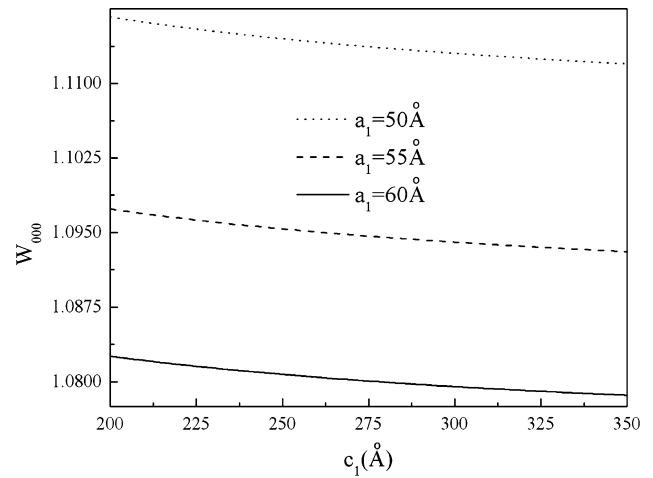

**Fig. 3** Dependences of the absorption threshold on the large semiaxis of the SPEQD at a fixed value of the small semiaxis

In the regime of intermediate SQ, the influence of the electron–hole Coulomb interaction is exhibited by means of the coefficients $\alpha$ and $\beta$ in formulas (12) to (14). Note that with the limiting transition $\alpha \to 0$, $\beta \to 0$, we arrive at the results of the regime of strong SQ.

In the regime of weak SQ, when the particle motion is determined by the Coulomb interaction and the contribution of SQ is a correction to it, as seen from formula (18), the families of equidistant levels caused by the SQ are disposed over each excitonic level. With increasing semiaxes, the equidistant levels are lowered and the interlevel distances decrease. Figures 4 and 5 show the dependences of the absorption edge on the semiaxes of the SPEQD. As is seen, in this case, the consideration of the Coulomb interaction leads to the decrease in the "effective" forbidden band width. In other words, the transitions are possible at lower frequencies of the incident light and, hence, in this case, the AE takes lesser values. From this, it follows that the consideration of the Coulomb interaction between an electron and hole leads to the shift of the AE to the long-wave region.

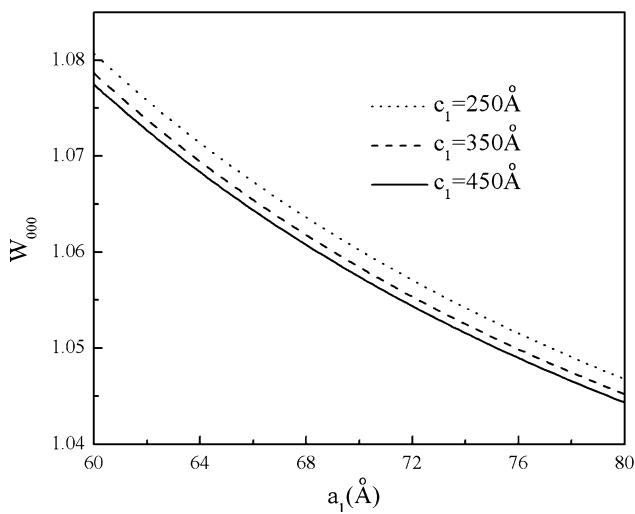

**Fig. 2** Dependences of the absorption threshold on the small semiaxis of the SPEQD at a fixed value of the large semiaxis

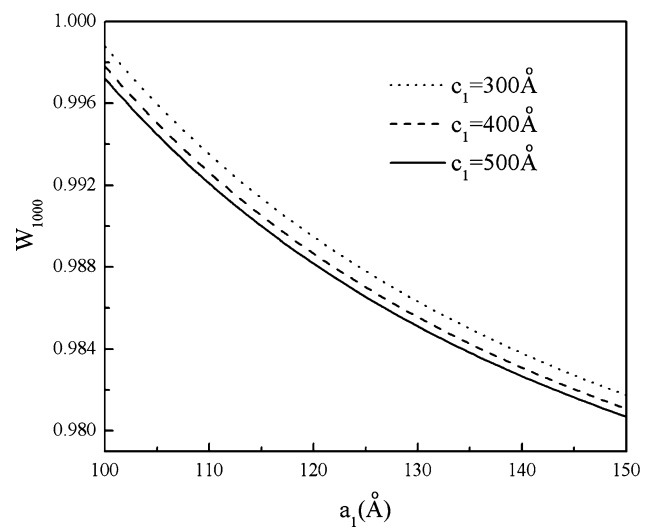

**Fig. 4** Dependences of the absorption threshold on the small semiaxis of the SPEQD at a fixed value of the large semiaxis

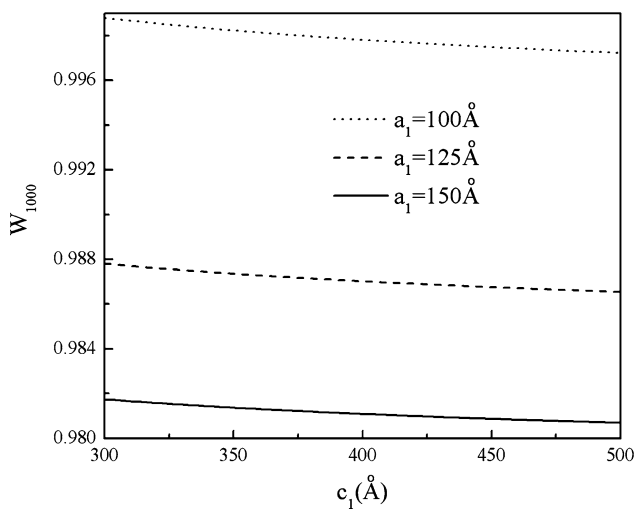

**Fig. 5** Dependences of the absorption threshold on the large semiaxis of the SPEQD at a fixed value of the small semiaxis

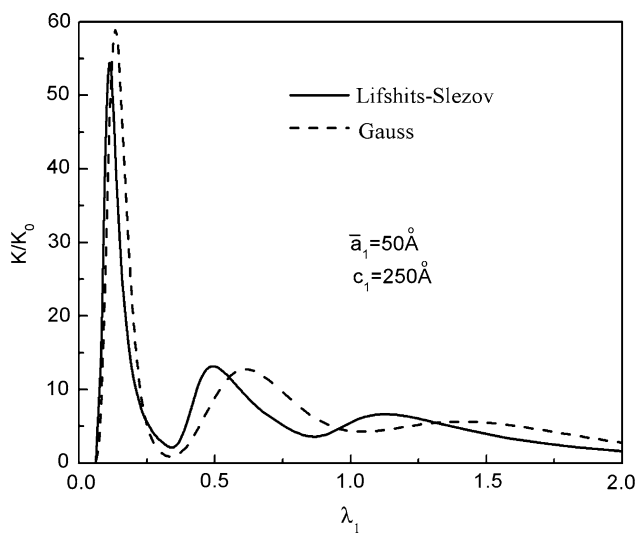

**Fig. 7** Dependence of absorption coefficient $K$ on the frequency of incident light, for the ensemble of SPEQDs for the strong SQ regime

Figure 6 shows the dependences of the electron ground state in a SPEQD, quantum wire, and cylindrical QD from *GaAs* (for equal values of the large semiaxis and cylinder height), respectively, on the small semiaxis, quantum wire radius, and radius of the cylindrical QD. As seen, the curve of the electron ground state energy in SPEQD is disposed higher, which is caused by the larger contribution of SQ into the particle energy as compared to other two cases.

Further, Fig. 7 illustrates the dependence of absorption coefficient $K$ on the frequency of incident light, for the ensemble of SPEQDs in strong SQ regime. As it mentioned above, instead of distinct absorption lines, account of size dispersion will give a series of fuzzy maximums. Note that both in the model of Gaussian and in the model of Lifshits–Slezov, QDs distributions a single distinctly expressed maximum of absorption is observed. When the light frequency is increased, the second weakly expressed maximum is seen. Further increase of the incident light frequency results in a fall of absorption coefficient.

Schematic diagrams for appropriate transitions, in which the absorption of light is present, are depicted in the Fig. 8 to understand in detail the process of absorption. From the comparison of diagram and Fig. 7, it is obvious that first clear expressed maximum corresponds to the $n = n' = 1$ transition family, and weak expressed picks are the result of the transitions between equidistant levels. The second weaker pick corresponds to the $n = n' = 2$ transition family. It is also obvious from Fig. 7 that the intensity corresponding to the above-mentioned family is weaker than the first maximum. This is the result of the small volume of the overlapping integral. The above-mentioned means, that the probability of transition is decreased.

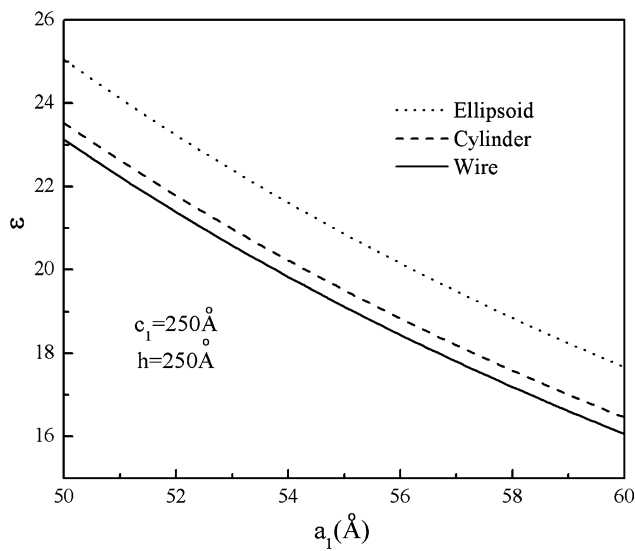

**Fig. 6** Dependences of the electron ground state in a SPEQD, quantum wire, and cylindrical QD, respectively, on the small semiaxis, quantum wire radius, and radius of the cylindrical QD

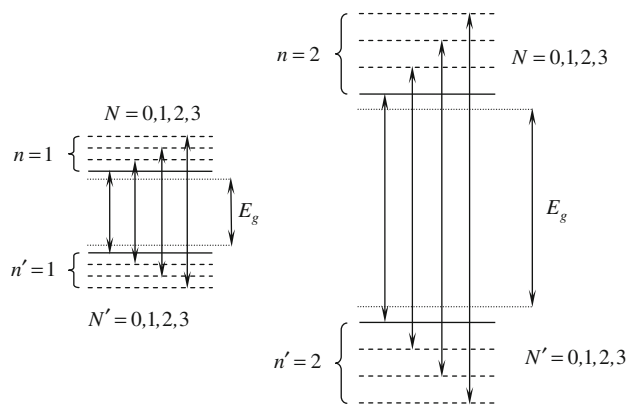

**Fig. 8** Schematic plot of corresponding interband transitions

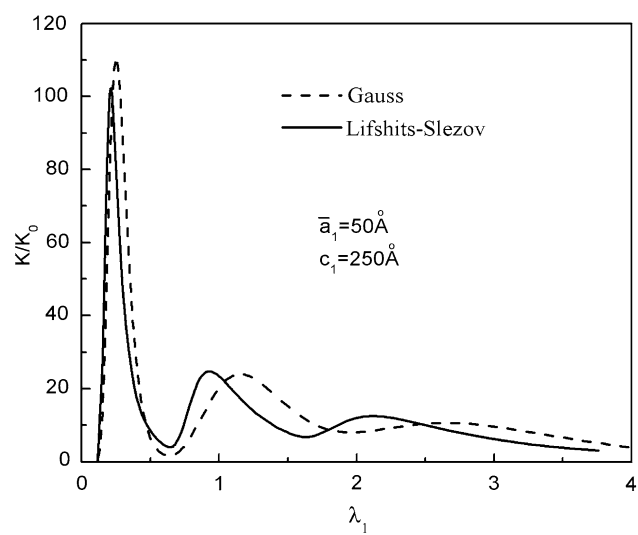

**Fig. 9** Dependence of absorption coefficient $K$ on the frequency of incident light, for the ensemble of SPEQDs for the weak SQ regime

Finally, Fig. 9 illustrates the dependence of absorption coefficient $K$ on the frequency of incident light, for the ensemble of SPEQDs in weak SQ regime. As can be seen from the picture taking into account the Coulomb interaction leads to the appearing of secondary well-expressed maximums of absorption. By the other word, the difference with the previous case has the quantitative character.

## Conclusion

In this work, we obtained that the electron energy is equidistant inside SPEQD in all three SQ regime cases. The impact of the dispersion of geometrical sizes for the QDs ensemble on direct light absorption is also investigated.

The SPEQDs, as more realistic nanostructures than quantum wires, have various commercial applications, in particular, in large two-dimensional focal plane arrays in the mid- and far infrared (M&FIR) region they have important applications in the fields of pollution detection, thermal imaging object location, and remote sensing as well as infrared imaging of astronomical objects.

These optimized quantum structures can be formed by direct epitaxial deposition using a self-assembling QDs technique, e.g., described in US Patent # 6541788 entitled "Mid infrared and near infrared light upconverter using self-assembled quantum dots" as well as by usage of MBE, MOCVD, or MOMBE deposition systems.

This theoretical investigation of SPEQDs can be effectively used for direct applications in photonics as background for simulation model. For further investigations, it is also important to develop a scheme for optimization of growth of SPEQDs needed for second harmonic generation.

**Acknowledgments** This work was carried out within the framework of the Armenian State Program "Semiconductor Nanoelectronics" and ANSEF Grant # PS NANO-1301, 2008.

## References

1. P. Harrison, *Quantum Wells, Wires and Dots, Theoretical and Computational Physics* (Wiley, NY, 2005)
2. D. Bastard, *Wave Mechanics Applied to Semiconductor Heterostructures* (Les editions de physique, Paris, 1989)
3. E.M. Kazaryan, S.G. Petrosyan, *Physical Principles of Semiconductor Nanoelectronics* (Izd. RAU, Yerevan, 2005)
4. M. Bayer, O. Stern, P. Hawrylak, S. Fafard, A. Forchel, Nature **405**, 923 (2000). doi:10.1038/35016020
5. C. Boze, C.K. Sarkar, Physica B **253**, 238 (1998). doi:10.1016/S0921-4526(98)00407-4
6. M. Califano, P. Harrison, J. Appl. Phys. **86**, 5054 (1999). doi:10.1063/1.371478
7. H. Chen, V. Apalkov, T. Chakraborty, Phys. Rev. B **75**, 193303 (2007). doi:10.1103/PhysRevB.75.193303
8. Z.M. Wang, K. Holmes, I.Y. Mazur, K.A. Ramsey, G.J. Salamo, Nanoscale Res. Lett. **1**, 57 (2006). doi:10.1007/s11671-006-9002-z
9. D.B. Hayrapetyan, J. Contemp. Phys. **42**, 293 (2007)
10. L.A. Juharyan, E.M. Kazaryan, L.S. Petrosyan, Solid State Commun. **139**, 537 (2006). doi:10.1016/j.ssc.2006.07.012
11. A.A. Tshantshapanyan, K.G. Dvoyan, E.M. Kazaryan, J. Mater. Sci: Mater. Electron. (2008). doi:10.1007/s10854-008-9753-7
12. K.G. Dvoyan, E.M. Kazaryan, Phys. Status Solidi. b **228**, 695 (2001). doi:10.1002/1521-3951(200112)228:3<695::AID-PSSB695>3.0.CO;2-P
13. K.G. Dvoyan, E.M. Kazaryan, L.S. Petrosyan, Physica E **28**, 333 (2005)
14. P. Maksym, T. Chakraborty, Phys. Rev. Lett. **65**, 108 (1990). doi:10.1103/PhysRevLett.65.108
15. D.B. Hayrapetyan, K.G. Dvoyan, E.M. Kazaryan, A.A. Tshantshapanyan, Nanoscale Res. Lett. **2**, 601 (2007). doi:10.1007/s11671-007-9079-z
16. V.M. Galitsky, B.M. Karnakov, *Kogan VI Practical Quantum Mechanics* (Nauka, Moscow, 1981)
17. A.I. Anselm, *Introduction to Semiconductors Theory* (Nauka, Moscow, 1978)
18. Al.L. Efros, A.L. Efros, Sov. Phys. Semicond. **16**, 772 (1982)
19. I.M. Lifshits, V.V. Slezov, Sov. Phys. JETP **35**, 479 (1958)
20. D. Leonard, M. Krishnamurthy, C.M. Reaves, S.P. Denbaars, P.M. Petroff, Appl. Phys. Lett. **63**, 23 (1993). doi:10.1063/1.110199